\documentclass[aip,cha,amsmath,amssymb,reprint,floatfix]{revtex4-1}

\usepackage{graphicx}% Include figure files
\usepackage{dcolumn}% Align table columns on decimal point
\usepackage{bm}% bold math
\usepackage[utf8]{inputenc}
\usepackage[T1]{fontenc}
\usepackage{mathptmx}
\usepackage{etoolbox}
\usepackage{hyperref}
\usepackage{xcolor}
\makeatletter
\def\@email#1#2{%
 \endgroup
 \patchcmd{\titleblock@produce}
  {\frontmatter@RRAPformat}
  {\frontmatter@RRAPformat{\produce@RRAP{*#1\href{mailto:#2}{#2}}}\frontmatter@RRAPformat}
  {}{}
}%
\makeatother

\begin{document}

\preprint{Chaos}

\title[Post-Disaster Resource Redistribution and Cooperation Evolution Based on Two-Layer Network Evolutionary Games]{Post-Disaster Resource Redistribution and Cooperation Evolution Based on Two-Layer Network Evolutionary Games}

\author{Yu Chen}
\affiliation{School of Mathematics and Statistics, Northwestern Polytechnical University, Xi'an, 710072, Shaanxi, China}
\affiliation{%
	Shaanxi Provincial Key Laboratory of Intelligent Game Theory and Information Processing at Higher Education Institutions, Xi'an, 710072, China}%

\author{Genjiu Xu}%
 \email{xugenjiu@nwpu.edu.cn.}
 \homepage{xugenjiu@nwpu.edu.cn.}
 \affiliation{School of Mathematics and Statistics, Northwestern Polytechnical University, Xi'an, 710072, Shaanxi, China}%
 \affiliation{Shenzhen Research Institute, Northwestern Polytechnical University, Shenzhen, 518057, Guangdong, China}%

\author{Sinan Feng}
 \affiliation{School of Mathematics and Statistics, Northwestern Polytechnical University, Xi'an, 710072, Shaanxi, China}
 \affiliation{%
 	Shaanxi Provincial Key Laboratory of Intelligent Game Theory and Information Processing at Higher Education Institutions, Xi'an, 710072, China}%

\author{Chaoqian Wang}
\affiliation{%
	School of Mathematics and Statistics, Nanjing University of Science and Technology, Nanjing 210094, China}%

\date{\today}% It is always \today, today,
             %  but any date may be explicitly specified

\begin{abstract}
In the aftermath of large-scale disasters, the scarcity of resources and the paralysis of infrastructure raise severe challenges to effective post-disaster recovery. Efficient coordination between shelters and victims plays a crucial role in building community resilience, yet the evolution of two-layer behavioral feedback between these two groups through network coupling remains insufficiently understood. Here, this study develops a two-layer network to capture the cross-layer coupling between shelters and victims. The upper layer uses a post-disaster emergency resource redistribution model within the framework of the public goods game, while the lower layer adopts a cooperative evolutionary game to describe internal victim interactions. Monte Carlo simulations on scale-free networks reveal threshold effects of incentives: moderate public goods enhancement and subsidies promote cooperation, whereas excessive incentives induce free-riding. In contrast, credible and well-executed punishment effectively suppresses defection. Targeted punishment of highly connected shelters significantly enhances cooperation under resource constraints. A comparative analysis using a network generated from the actual coordinates of Beijing shelters confirms the model's generality and practical applicability. The findings highlight the importance of calibrated incentives, enforceable sanctions, and structural targeting in fostering robust cooperation across organizational and individual levels in post-disaster environments.
\end{abstract}

\maketitle

\begin{quotation}
This study develops a structured two-layer network framework to facilitate coordinated cooperation between shelters and victims under post-disaster resource scarcity, thereby mitigating disaster impacts and accelerating recovery. The upper layer adopts a public goods-based emergency resource redistribution model to characterize continuous cooperative inputs among shelters, while the lower layer uses a cooperative evolutionary game to represent binary interactions among victims. This coupled framework captures asymmetric cross-layer feedback between organizational and individual behaviors. Simulations show that calibrated incentives and credible sanctions sustain cooperation, whereas excessive incentives induce free-riding. Targeted sanctions applied to highly interconnected shelters further enhance system-wide cooperation at lower cost. A comparative analysis using a network constructed from the actual geographical coordinates of shelters in Beijing confirms the generality and applicability of the model. These findings offer practical insights for policymakers seeking to design effective post-disaster management strategies that balance top-down coordination with grassroots collaboration.
\end{quotation}

\section{Introduction}
Sudden disasters such as earthquakes, hurricanes and large-scale floods often lead to a serious shortage of key resources, paralysis of infrastructure, and widespread panic \cite{Comfort2007,Drabek2012,May2021}. These interruptions have seriously damaged the efficiency of emergency response and post-disaster recovery. Effective early post-disaster response is crucial to reducing casualties but is often hindered by communication breakdowns and limited resources \cite{Wang2024,Aalami2018,Kovacs2009,Liu2020,Akter2019}. During this critical period, shelters serve not only as gathering points but also as primary organizational units for allocating and redistributing emergency resources. Victims in these shelters can cooperate spontaneously to improve their collective survival opportunities \cite{hanson2008catastrophe,shoaf2000public}. However, sustaining cooperation under scarcity and survival pressure remains a major challenge \cite{Aldrich2012,Nakagawa2004}. 
 
Previous studies have focused on two main areas. The first line of research focuses on optimizing emergency resource allocation using mathematical models that incorporate efficiency, cost, and fairness criteria \cite{Mete2010,Ozdamar2015,Wang2024}. Early research focused on efficiency as a primary goal, often measured by delivery time \cite{Altay2013,Berkoune2012}. Other research emphasizes cost \cite{Arrubla2014,Minas2015,Ozdamar2004,Yi2007,Han2021} or fairness \cite{Erbeyoglu2020,Kovacs2009}. Some studies have integrated multiple criteria to address trade-offs in real rescue scenarios \cite{Salmeron2010,Sheu2007,Wang2021}.

The second is to use evolutionary game theory to explore how cooperation arises and stabilizes in the face of adversity. The tension between individual and collective welfare creates a dilemma of cooperation \cite{Orbell1991,Orbell1984,Seale2006}. Mechanisms such as kinship selection \cite{Dawkins2016,West2002}, network reciprocity \cite{Traulsen2006,Lieberman2005}, persistence \cite{Wang2023Evolution,wang2023inertia,wang2023conflict,wang2023greediness}, emotion \cite{Szolnoki2013}, memory \cite{Wang2006,Huang2024}, and trust \cite{Abbass2015,Wang2024Evolution,Wang2025Inter} provide theoretical support for cooperation strategies under resource pressure.

Despite these advances, few studies have explored the dynamic interactions between organizational structures (shelters) and individual behaviors (victims). Most models either assume centralized control by relief agencies or examine isolated group behavior without accounting for cross-layer feedback. To address this question, this study develops a two-layer network model that captures the cross-layer feedback between shelters and victims. 

The rest of the study is organized as follows. The literature review section outlines prior research on emergency resource allocation and cooperation mechanisms. The methodology section introduces the two-layer network model and simulation setup. Then the simulation results and case analysis are presented. The final part presents the discussions and conclusion.

\section{Literature Review}
\subsection{Emergency resource management and allocation}
Most studies on emergency resource allocation have constructed optimization models focusing on efficiency, cost, fairness \cite{Wang2024}. Efficiency is usually prioritized by minimizing delivery time \cite{Altay2013,Berkoune2012}, and \citet{Yan2009,Wex2014} developed a time minimization model for material allocation. In addition, cost minimization is also a key issue. \citet{Barbarosoglu2002} proposed a two-stage stochastic programming model for emergency resource allocation to minimize the total cost. \citet{Cotes2019} developed a model to locate and allocate humanitarian emergency facilities to respond to floods. In addition to efficiency and cost, some scholars are gradually guided by distributive justice. \citet{Ahmadi2022} found that ignoring the fair standards of emergency resource allocation may lead to victim dissatisfaction and social unrest.  \citet{Huang2019} found that the key to effectively reducing casualties and losses lies in the fair allocation of emergency resources. \citet{Wang2019} measures the fairness of distribution by minimizing the loss of non-public utilities resulting from resource shortages. To address the practical trade-off problem, some works integrate multiple objectives, such as efficiency and cost \cite{Fu2018}, or efficiency and fairness \cite{li2023coordinated}. A comprehensive multi-objective approach is also proposed, taking into account efficiency, cost and fairness \cite{Tzeng2007,Bozorgi2016}. 

\subsection{Evolutionary mechanisms of cooperation}
In a resource-scarce environment, victims may help each other or resort to hoarding and selfish behavior. Evolution game theory provides a basis for understanding cooperation under such adversity \cite{Hamilton1964,Nowak2006}. Scholars have proposed a variety of mechanisms based on the public goods game model, such as reward mechanism \cite{Szolnoki2011,Xie2023}, punishment mechanism \cite{Wang2021Tax}, and voluntary selection mechanism \cite{Hauert2002}. In the governance of the real world, institutions play the role of ``punishing evil and promoting good'', encouraging cooperation without direct returns \cite{Sigmund2001,Andreoni2003,Sigmund2007,Fehr2002,wang2024evolutionary}. Group-level incentives have shown notable effectiveness \cite{Bahbouhi2024}, while studies on reputation systems offer further insights into cooperation dynamics \cite{Wang2017,Lv2024}.

Multilayer and interdependent networks have become a key framework for understanding cooperation beyond single-layer structures. Early studies showed that optimal interdependence, interdependent network reciprocity, and biased cross-layer utility functions can substantially enhance cooperation \cite{Wang2013SciRep2470,Wang2013SciRep1183,Wang2012EPL48001}. A comprehensive synthesis of these mechanisms is provided in the multilayer evolutionary games colloquium \cite{Wang2015EPJB124}. More recent studies show that coupling group selection with network reciprocity or linking layers through hyperbolic structures can further promote cooperation \cite{Shi2022AMC126835,Duh2023EPL62002}. Building on these foundations, applied studies have explored how incentives, social influence, and structural coupling shape cooperation in multilayer or interdependent systems \cite{Jiang2025,Su2023,Zhu2025,Wang2023Public}. These studies show that inter-layer interactions and structural heterogeneity substantially influence cooperative dynamics.

In summary, scholars have discussed the allocation of emergency resources, cooperation mechanisms, and network structures; however, they tend to ignore behavioral feedback between shelters and victims, rely on binary strategies, or ignore asymmetric cross-layer interactions. To fill these gaps, this study constructs a two-layer network model that couples shelter-level resource redistribution with victim-level cooperation, incorporating continuous strategies, government interventions, and cross-layer feedback. Simulation experiments examine the effects of enhancement factors, subsidies, and punishments. A case study based on shelter coordinates in Beijing validates the model and offers insights for strengthening cooperation in the early post-disaster stage.

\section{Methodology}

\subsection{Model design}
\begin{figure}
	\centering
	\includegraphics[width=0.6\columnwidth]{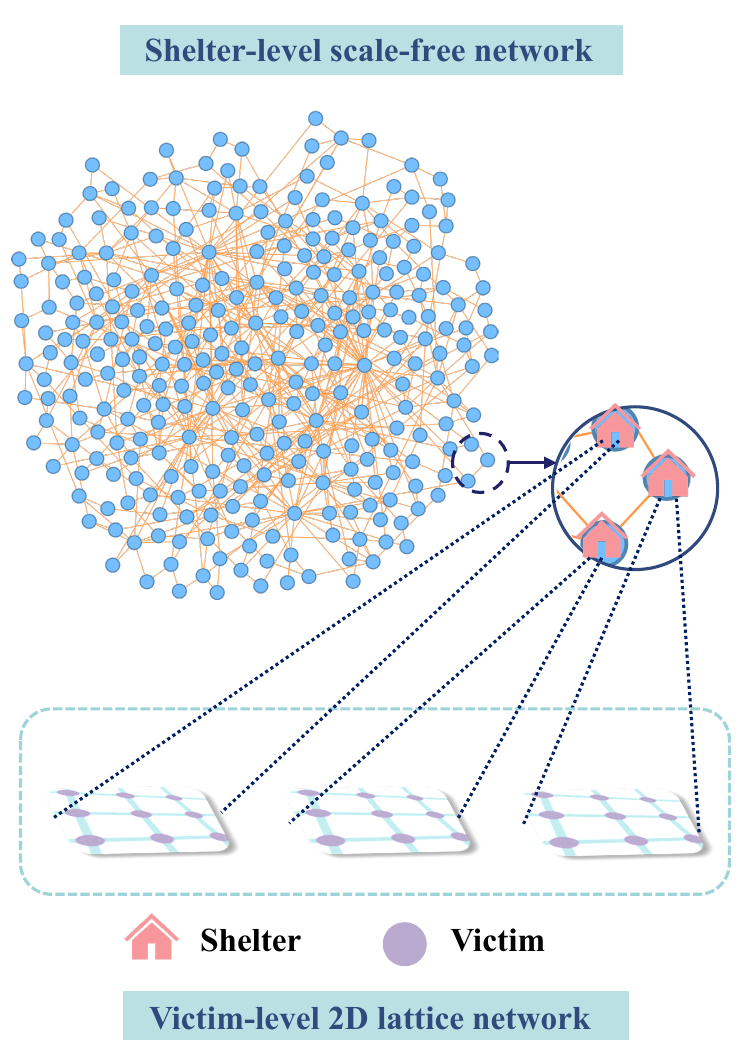}
	\caption{Schematic of the coupled two-layer ``shelter--victim'' network.}
	\label{two_layer_network}
\end{figure}

Post-disaster shelter interactions are modeled as a scale-free network, while the victims within each shelter are distributed on an $L\times L$ two-dimensional lattice. As shown in Fig.~\ref{two_layer_network}, we construct a two-layer model that couples a continuous public goods game among shelters (upper layer) with an evolutionary cooperation game among victims within each shelter (lower layer). The set of participants $N=\{1,\dots,n\}$ represents all shelters in the upper-layer network. The set of participants $V_i = \{v_{i,1},\dots,v_{i,m_i}\}, \quad i=1,\dots,n$ denotes all victims within each shelter.

On the upper-layer network, we consider $n$ shelters. 
To alleviate resource shortages and panic, shelters participate in a continuous-strategy public goods game. For each shelter $i$, the ideal resource amount is denoted as $x_i^t \in [a_1,a_2]$, and the current available resource is $y_i^t \in [a_3,a_4]$. At each time step $t$, the strategy set of participant $i$ is $S \subseteq [0, y_i^t]$, the strategy \( s_i \) of participant \( i \in N \) is defined as
\begin{equation}
	s_i = \theta_i^t y_i^t,\quad \theta_i^t \in [0,1], \tag{1}
\end{equation}
where $\theta_i^t$ is the input coefficient of participant $i$. As $\theta_i^t$ approaches 0, participant $i$ contributes less and exhibits a stronger tendency to defect. Each participant $i$ and its neighbors form a group, denoted as $G_i$.

Table~\ref{tb_symbol} defines the sets, parameters, variables, and dynamics used in the model across both layers. 

\begin{table*}
\small
\caption{\label{tb_symbol}Notation for the two-layer shelter--victim cooperation model.}
\begin{ruledtabular}
\begin{tabular}{p{0.16\textwidth}p{0.78\textwidth}}
\multicolumn{2}{l}{\textbf{Sets and indices}}\\
$N$      & Set of shelters, indexed by $i\in N$.\\
$V_i$    & Set of victims in shelter $i$, indexed by $j\in V_i$.\\
$G_i$      & Group centered at shelter $i$, consisting of $i$ and its network neighbors.\\
$t$      & Time step in the evolutionary process.\\[2pt]

\multicolumn{2}{l}{\textbf{Parameters}}\\
$m_i$          & Number of victims in shelter $i$.\\
$r\ge 1$       & Enhancement factor of public goods.\\
$\sigma\in[0,1]$ & Government subsidy rate for contributions.\\
$\gamma_1$     & The coefficient of marginal impact of the shelter punishment.\\
$\gamma^0$   & The coefficient of victim-baseline punishment.\\
$\alpha$       & The coefficient of punishment execution status.\\
$\kappa$       & Cooperation cost for each refugee.\\
$\beta$        & Cooperation threshold of victims that determines shelter payoff regime.\\
$k$            & Selection intensity in the Fermi rule (strategy update strength).\\
$[a_1,a_2]$    & Range of ideal shelter resources.\\
$[a_3,a_4]$    & Range of actual available shelter resources.\\[2pt]

\multicolumn{2}{l}{\textbf{Variables (upper layer: shelters)}}\\
$x_i^t$              & Ideal resource amount of shelter $i$ at time $t$.\\
$y_i^t$              & Available resource amount of shelter $i$ at time $t$.\\
$S$&Strategy set of shelters.\\
$s_i=\theta_i^t y_i^t$ & Contribution of shelter $i$, proportional to its available resources.\\
$\theta_i^t\in[0,1]$   & Contribution ratio of shelter $i$, with smaller values indicating stronger defection.\\
$\pi_G^t$            & Total payoff of group $G$ at time $t$.\\
$\pi_i^t$            & Payoff of shelter $i$ at time $t$.\\[2pt]

\multicolumn{2}{l}{\textbf{Variables (lower layer: victims)}}\\
$A_{i,j}\in\{C,D\}$  & Strategy of refugee $j$ in shelter $i$, where $C=1$ (cooperate), $D=0$ (defect).\\
$\mu_i^t$            & Proportion of cooperating victims in shelter $i$ at time $t$.\\
$R_i^t=\pi_i^t/m_i$    & Per capita payoff of victims in shelter $i$ at time $t$.\\
$\gamma_2^t=\gamma^0+\alpha(1-\theta_i^t)$ & Punishment intensity for defecting victims in shelter $i$.\\
$P_{i,j}$            & Payoff of refugee $j$ in shelter $i$.\\[2pt]

\multicolumn{2}{l}{\textbf{Dynamic variables}}\\
$W(s_i\rightarrow s_j)$  & Probability that shelter $i$ imitates the strategy of shelter $j$ (Fermi update).\\
$W(A_{i,n}\rightarrow A_{i,m})$ & Probability that refugee $v_{i,n}$ imitates the strategy of neighbor $v_{i,m}$.\\
\end{tabular}
\end{ruledtabular}
\end{table*}

The total utility of the group is given by
\begin{equation}
	\pi_G^t = r\sum_{i\in G}s_i = r\sum_{i=1}^g \theta_i^t\,y_i^t, \tag{2}
\end{equation}
where $r \geq 1$ reflects the enhancement factor, indicating the extent to which cooperation amplifies the original resources' utility.

The payoff function of shelter $i$ is defined as
\begin{equation}\label{eq_pi}
	\pi_i^t =
	\begin{cases}
		y_i^t - s_i + \dfrac{y_i^t}{\sum_{j\in G}y_j^t}\pi_G^t + \sigma\,s_i, & \mu_i^t \geq \beta, \\[8pt]
		y_i^t - s_i + \dfrac{y_i^t}{\sum_{j\in G}y_j^t}\pi_G^t - \gamma_1 y_i^t, & 0 \leq \mu_i^t < \beta,
	\end{cases} \tag{3}
\end{equation}
where $\sigma \in [0,1]$ denotes government subsidies and $\gamma_1$ represents the punishment intensity. The payoff type is further determined by the cooperation level of victims within each shelter.

At each Monte Carlo step, two distinct participants \(i,j\in N\) are drawn uniformly at random (they need not be neighbors). Participant \(i\) forms a group with all of its network neighbors, and \(j\) likewise forms a group with its own neighbors. We compute each player's payoff, \(\pi_{i}^t\) and \(\pi_{j}^t\), from their respective groups. Then \(i\) updates its strategy by imitating \(j\) with the probability,
\begin{equation}\label{eq_W}
	W(s_i\rightarrow s_j)
	= \frac{1}{1+\exp\bigl[-k(\pi_j^t-\pi_i^t)\bigr]}. \tag{4}
\end{equation}
Here $k$ is the selection intensity; when $k$ approaches 0, participants imitate randomly.

\medskip

On the lower-layer network, multiple independent lattice networks represent the victim groups. Each node corresponds to a victim who adopts a binary strategy $A_{i,j}\in \{C,D\}$, where $C=1$ (cooperate) and $D=0$ (defect). At time $t$, shelter $i$ obtains a total payoff $\pi_i^t$ from the upper-layer continuous-strategy public goods game. The cooperation level of victims in shelter $i$ is defined as
\begin{equation}
	\mu_i^t = \frac{1}{m_i}\sum_{j\in V_i} A_{i,j}. \tag{5}
\end{equation}

Every round the basic payoff of each victim is
\begin{equation}
	R_i^t = \frac{\pi_i^t}{m_i}, \tag{6}
\end{equation}
where each cooperator pays a cooperation cost $\kappa$, and defectors are punished as
\begin{equation}
	\gamma_2^t = \gamma^0 + \alpha \bigl(1-\theta_i^t\bigr), \tag{7}
\end{equation}
where $\gamma^0$ is the victim-baseline punishment, and $\alpha$ is the victims-punishment execution coefficient, which intensifies when the local cooperation level is low.

Thus, the payoff matrix of the evolutionary game among victims is
\begin{equation}
	\begin{array}{c|cc}
		& C & D \\ \hline
		C & \displaystyle{R_i^t - \frac{\kappa}{2}} & R_i^t - \kappa \\[4pt]
		D & R_i^t - \gamma_2^t & 0 
	\end{array} . \tag{8}
\end{equation}

Similar to the upper-layer network, victim $v_{i,n}$ imitates the strategy of neighbor $v_{i,m}$ with probability following the Fermi rule:
\begin{equation}
	W\bigl(A_{i,n}\rightarrow A_{i,m}\bigr)
	= \frac{1}{1+\exp\bigl[-k\bigl(P_{i,m}-P_{i,n}\bigr)\bigr]}, \tag{9}
\end{equation}
where $P_{i,n}$ and $P_{i,m}$ denote the respective payoffs.

\subsection{Solution method}
The proposed two-layer model integrates a continuous-strategy public goods game among shelters with a cooperative evolutionary game among victims, forming a high-dimensional nonlinear coupled network. In the upper layer, payoffs depend on individual strategy $s_i = \theta_i^t y_i^t$, group payoff $\pi_G^t$ and the cooperation degree $\mu_i^t$ [Eq.~(\ref{eq_pi})]. Shelters' strategies are updated by the Fermi rule [Eq.~(\ref{eq_W})]. Coupling and threshold feedback induce complex, nonsmooth dynamics, hindering analytical solutions. 

Previous studies such as Szolnoki \textit{et al.} \cite{Szolnoki2011} and \citet{Wang2019} employed Monte Carlo simulations to examine evolutionary dynamics. In this study, local equilibria are derived for each layer under fixed conditions, and overall dynamics are subsequently examined via numerical simulations.  

Given fixed neighbor strategies $\{s_j, j\in G, j\neq i\}$ and cooperation level $\mu_i^t$, the local optimal condition $\partial \pi_i / \partial s_i = 0$ yields two cases.

When $\mu_i^t \ge \beta$,
\begin{equation}
	\theta_i^* = \frac{1 - \dfrac{y_i^t}{\sum_{j\in G} y_j^t} r}{\sigma}, \quad s_i^* = \theta_i^* y_i^t. \tag{10}
\end{equation}

When $\mu_i^t < \beta$,
\begin{equation}
	\theta_i^* = 1 - \frac{y_i^t}{\sum_{j\in G} y_j^t} r, \quad s_i^* = \theta_i^* y_i^t, \tag{11}
\end{equation}
and the feasible region requires $s_i^* \in [0, y_i^t]$.

For fixed upper-layer strategy $s_i$, the evolution of victim cooperation in shelter $i$ is approximated under weak selection by the replicator dynamics,
\begin{equation}
	\dot \mu_i \approx k\, \mu_i (1-\mu_i) (\Pi_C - \Pi_D), \tag{12}
\end{equation}
with payoff difference,
\begin{equation}
	\Pi_C - \Pi_D = (R_i^t - \kappa) + \mu_i \left(\frac{\kappa}{2} + \gamma_2^t - R_i^t\right), \tag{13}
\end{equation}
where $R_i^t = \pi_i^t/m_i$ and $\gamma_2^t = \gamma^0 + \alpha (1-\theta_i^t)$.

The fixed point condition $\Pi_C - \Pi_D = 0$ gives the local equilibrium cooperation level,
\begin{equation}
	\mu_i^* = \frac{\kappa - R_i^t}{\dfrac{\kappa}{2} + \gamma_2^t - R_i^t} \in [0,1], \tag{14}
\end{equation}
representing the stable cooperation state under given upper-layer strategies.

\subsection{Simulation study}
Monte Carlo simulations examine how $s_i$ and $\mu_i$ evolve under varying parameters, using two upper-layer networks to test model enforceability and generality.

\textbf{BA network} (Barabási--Albert scale-free network \cite{Albert2002}): a theoretical network generated via preferential attachment.

\textbf{BJ network} (Beijing-based spatial network): a real-world network was constructed from the geographic coordinates of Beijing shelters, based on data provided by the Beijing Municipal Emergency Management Bureau \cite{beijing2025}.

Numerical simulations study how the enhancement factor, government subsidies, and punitive measures affect the cooperative behavior between shelters and victims in the case of scarce post-disaster resources. The model captures macro--micro interactions through structural coupling between scale-free shelter networks and lattice-based victim groups. By varying these institutional parameters, the simulation explores governance regimes ranging from a ``completely free game'' to a ``strongly regulated game.'' Cross-layer cooperation dynamics follow the update sequence outlined in the supplementary material, which defines the feedback process used in all simulations.

\section{Simulation Results}
\subsection{Effects of enhancement factor and government intervention} 
\begin{figure}
  \centering
	\includegraphics [width=\columnwidth]{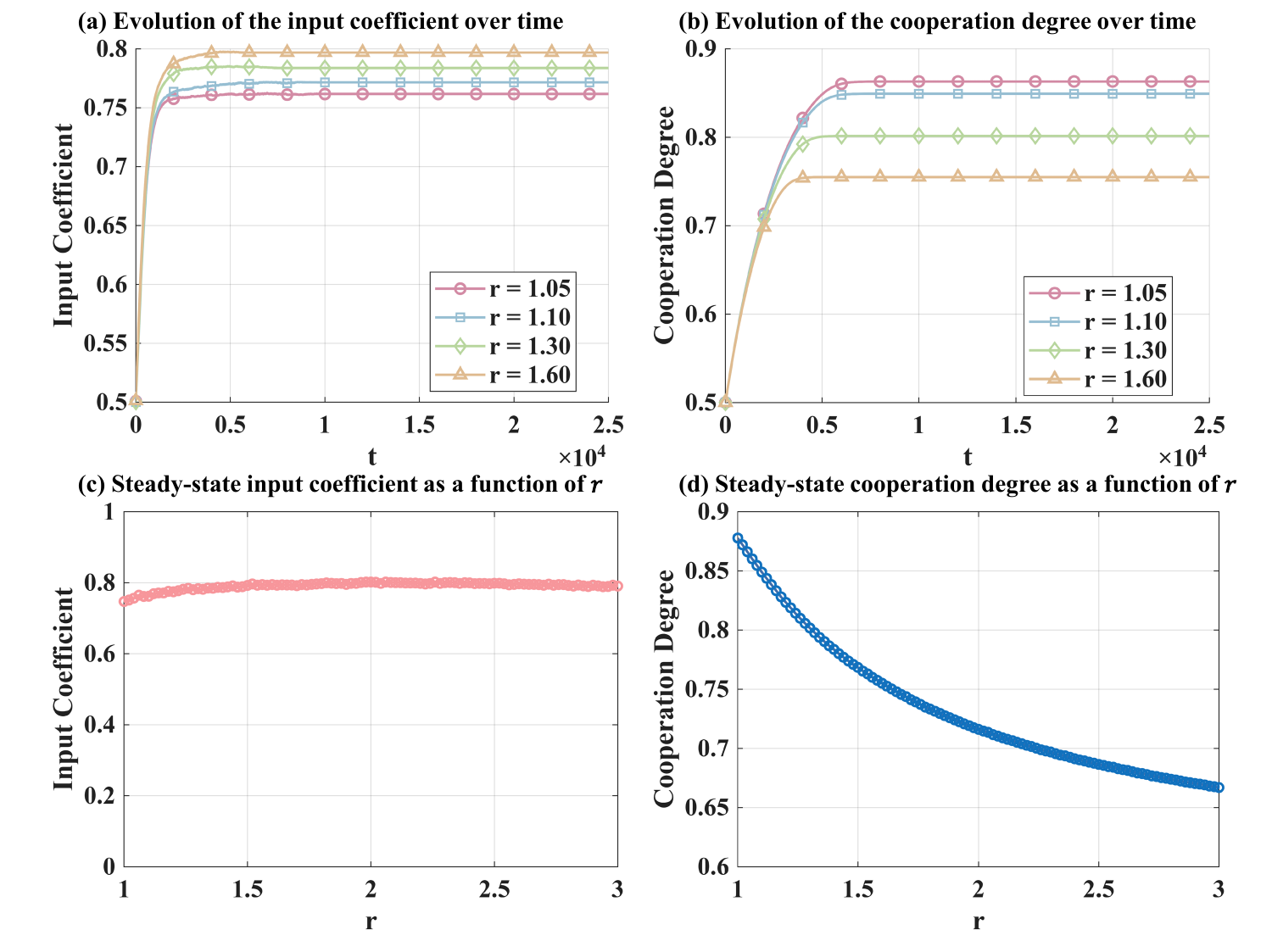}  
	\caption{Moderate enhancement promotes cooperation, excessive $r$ reduces it. (a)–(b) Evolution of $\theta(t)$, $\mu(t)$ under \(r \in \{1.05,\,1.10,\,1.30,\,1.60\}\). Moderate values accelerate convergence and yield higher cooperation, whereas large $r$ weakens incentives. (c)–(d) Steady-state of $\theta(t)$, $\mu(t)$ over $r \in [1.0, 3.0]$, showing a non-monotonic trend: shelter input rises slightly with $r$, while victim cooperation declines monotonically. Each result from 256 runs over 25 000 steps, with steady-state values averaged over the final 1000 steps. Parameters: $\sigma=0.40$, $\gamma_1=0.50$, $\alpha=0.50$, $\gamma^0=0.90$, and  $\beta=0.50$.
	}  
	\label{BA_r} 
\end{figure}

To assess external interventions in the post-disaster two-layer network, this study examines how varying the enhancement factor, subsidies, and punishment strength affects cooperation at both shelter and victim levels. As shown in Fig.~\ref{BA_r}, as the enhancement factor \(r\) increases, it initially promotes the upper-layer average input \(\theta\), but eventually weakens it, while the lower-layer cooperation degree \(\mu\) declines steadily. The findings indicate that moderate \(r\) enhances returns and encourages cooperation, but excessive \(r\) reduces marginal benefits, leading shelters to cut inputs and victims to free-ride. 

\begin{figure}  
	\centering  
	\includegraphics [width=\columnwidth]{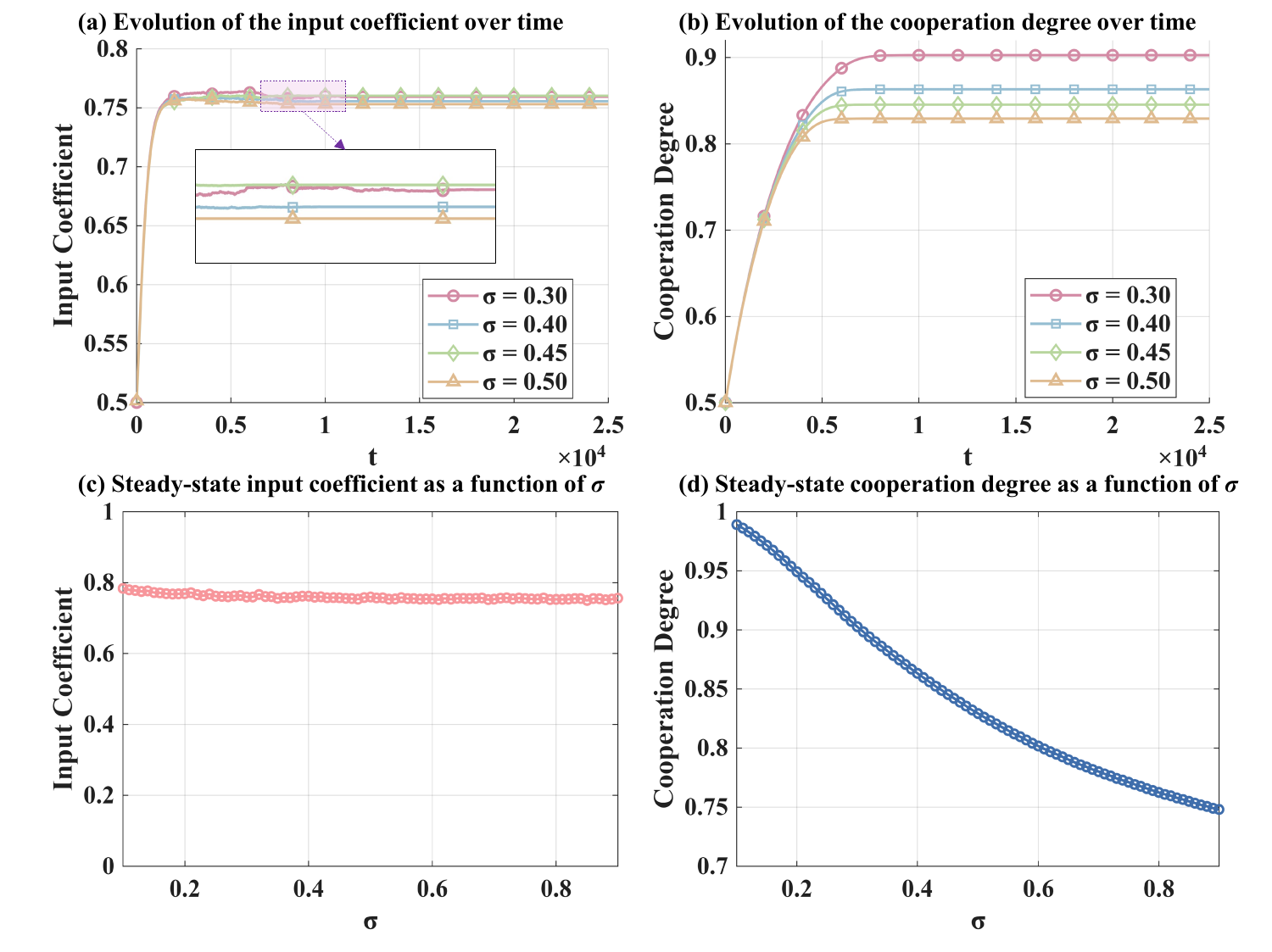}  
	\caption{Higher subsidy \(\sigma\) weakens lower-layer cooperation. 
		(a)–(b) Evolution of $\theta(t)$, $\mu(t)$ under \(\sigma \in \{0.30,\,0.40,\,0.45,\,0.50\}\). Shelters respond weakly to subsidies, but victims become less cooperative as $\sigma$ increases. (c)–(d) Steady-state of $\theta(t)$, $\mu(t)$ over range \([0,\,1)\). 
		Default parameters, \(r=1.30\), the simulation settings unchanged.} 
        \label{BA_sigma}
\end{figure}

As shown in Fig.~\ref{BA_sigma}, increasing government subsidies reduces the two-layer cooperation, which indicates that excessive rewards inhibit intrinsic cooperation motivation. When subsidies become too generous, shelters receive a steady payoff, allowing victims to benefit without cooperating and undermining their motivation to do so. Similar to the effect of \(r\), moderate subsidies promote cooperation, while excessive ones undermine it. Effective incentives require differentiation by shelter conditions or levels of internal victim cooperation, rather than being applied uniformly across all shelters.

As shown in Fig.~\ref{BA_P}, the victim-baseline punishment coefficient \(\gamma^0\) plays a decisive role in promoting cooperation, triggering a transition from low to high cooperation when exceeding 0.9. The execution coefficient \(\alpha\) enhances this effect by ensuring execution, while the marginal impact of the shelter punishment coefficient \(\gamma_1\) is limited and is mainly supplementary under weak cooperation. These results indicate that stable cooperation in a post-disaster environment requires strong and effectively implemented punishment.

Simulation results indicate that while moderate incentives promote cooperation, excessive ones weaken it, and only coordinated, enforceable punishment combinations can effectively sustain cross-layer cooperation.

\subsection{The influence of the combination and application of punishment on cooperation} 
To clarify the role of punishment, this section examines punishment parameter combinations and spatial applications, including random selection and structure-based diffusion, that affect cooperation and cross-layer feedback in two-layer networks.

\begin{figure}  
	\centering  
	\includegraphics [width=\columnwidth]{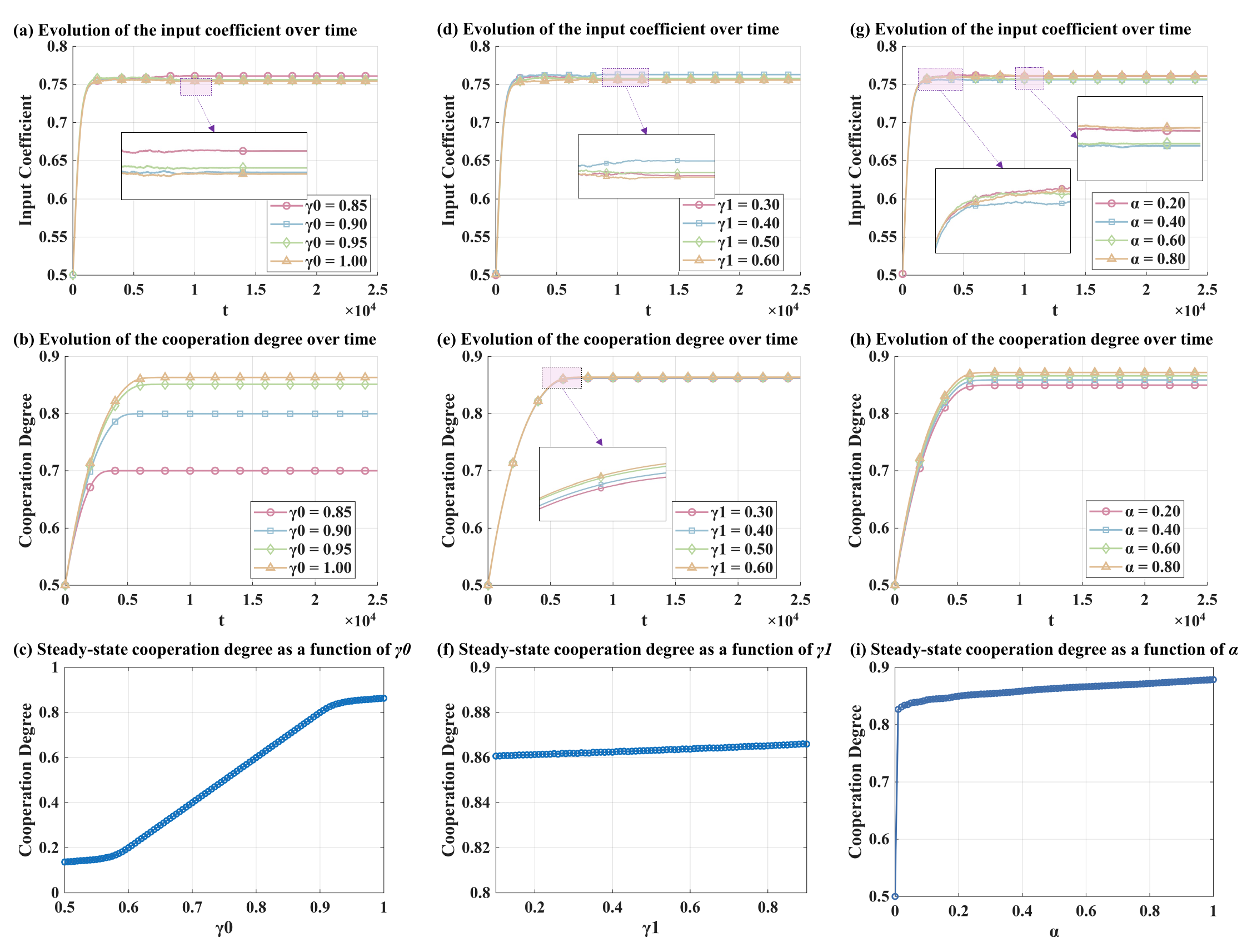}  
	\caption{\textbf{Punishment effect: \(\gamma^0\) dominates, \(\alpha\) has a secondary effect, and \(\gamma_1\) shows a relatively weak impact.} 
		Rows 1 and 2: the evolution of $\theta(t)$ and $\mu(t)$ as each parameter varies independently.
		Row 3: steady-state of $\mu(t)$ for each parameter.  
		Default parameters, \(r=1.30\), \(\sigma=0.40\), the simulation settings unchanged.}  
    \label{BA_P}
\end{figure}

\begin{figure}
	\centering  
	\includegraphics [width=\columnwidth]{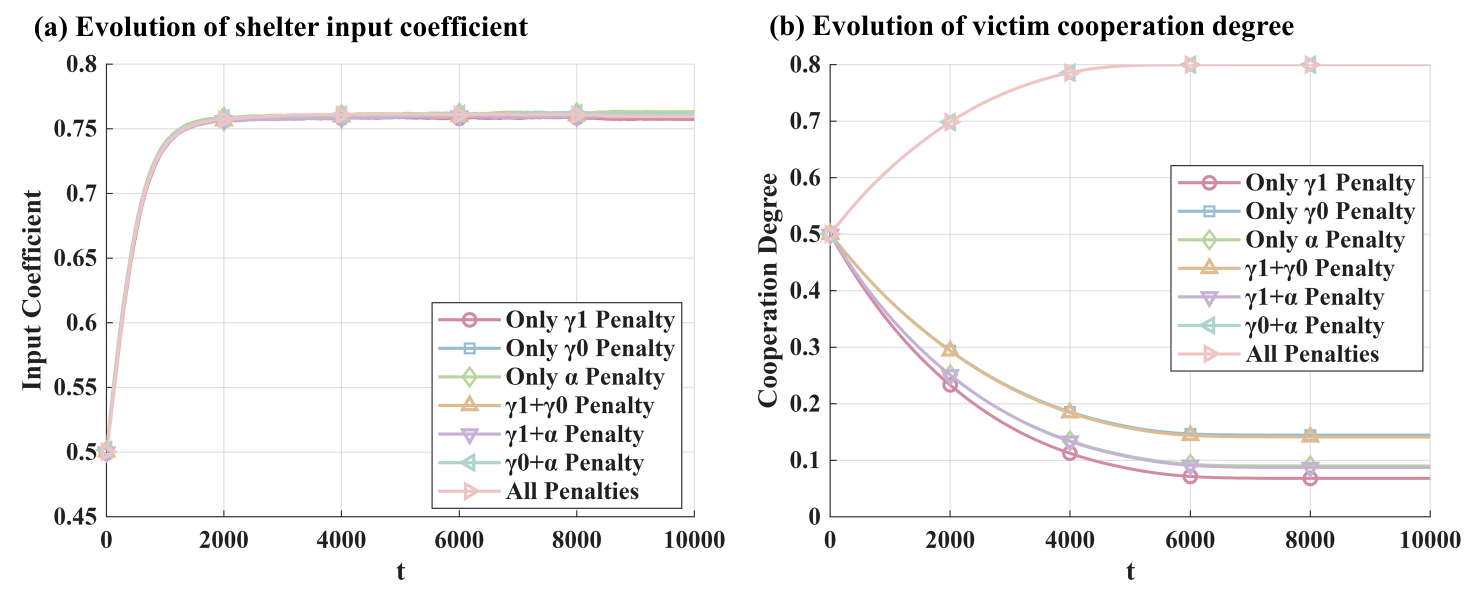}  
	\caption{\textbf{Joint punishment maximizes cooperation.} 
	(a)–(b) Evolution of \(\theta(t)\) and \(\mu(t)\) under seven punishment configurations. Single-parameter punishment is insufficient, and only full penalty combination sustains high cooperation. Default parameters, the simulation settings unchanged.}  
	\label{5P}  
\end{figure}

The results in Fig.~\ref{5P} indicate that different punishments exert limited influence on shelter input but markedly affect victim cooperation. Single measures lead to low cooperation, while adding the victim-baseline punishment coefficient $\gamma^0$ significantly improves it. Maximum cooperation is achieved when all three penalties are combined. Mechanistically, $\gamma^0$ suppresses defection, $\alpha$ ensures execution, and $\gamma_1$ offers supplementary constraints. These findings, consistent with Fig.~\ref{BA_P}, confirm that only strong and enforceable punishment ensures stable victim cooperation after disasters.

\begin{figure}  
	\centering  
	\includegraphics [width=\columnwidth]{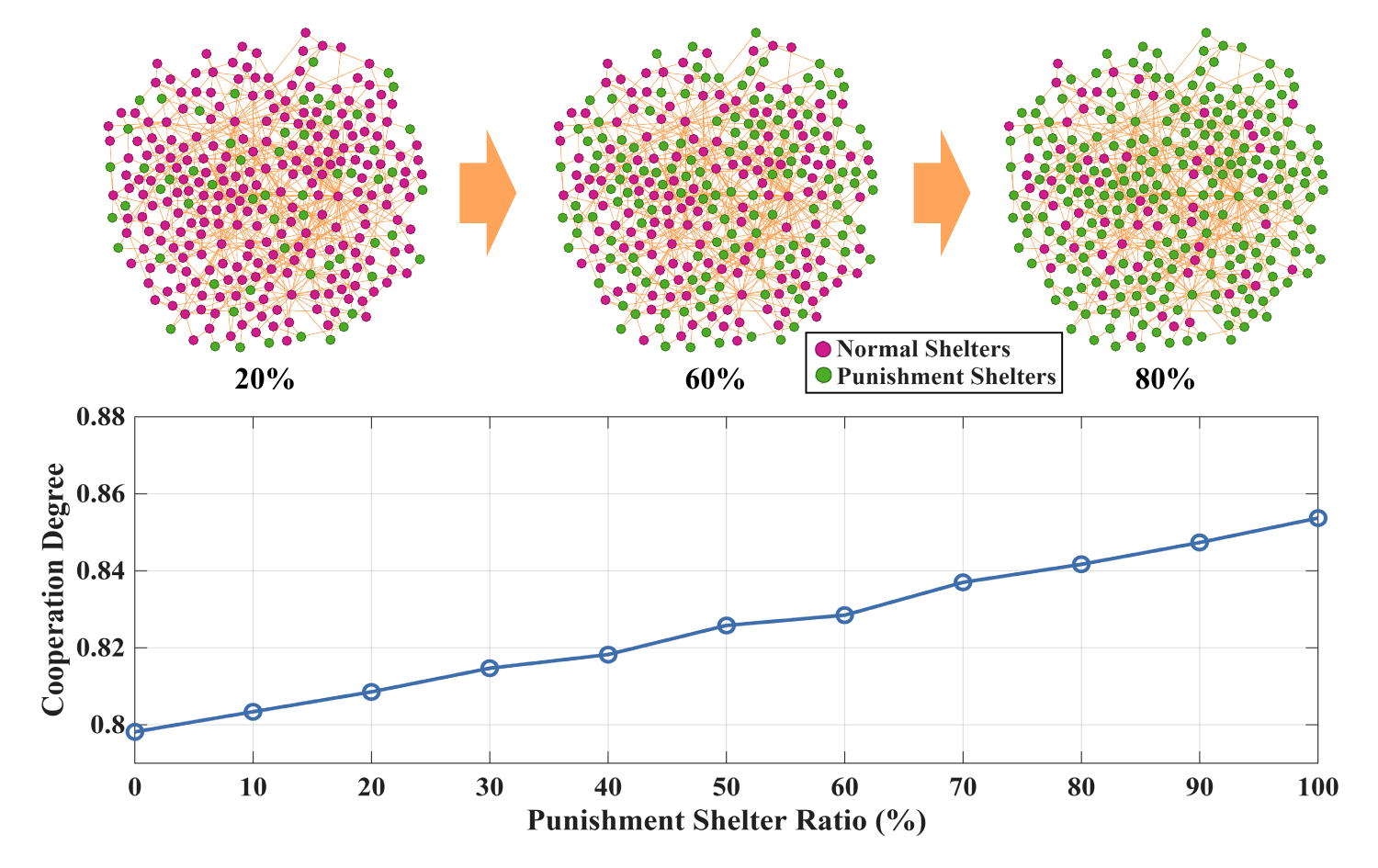}  
	\caption{Increasing the proportion of punished shelters improves cooperation almost linearly. Upper: \(20\%\), \(60\%\), and \(80\%\) of shelters are randomly selected to apply punishment. Lower: cooperation degree \(\mu(t)\) with punishment ratio, showing a near-linear improvement. Default parameters, the simulation settings unchanged.}  
	\label{6PR}  
\end{figure}

Furthermore, Fig.~\ref{6PR} indicates that random deployment of punishment in upper-layer shelters leads to an almost linear rise in victim cooperation as coverage increases. This effect is due to cross-layer transmission of punishment signals via income redistribution and strategic imitation, which indirectly restrains the behavior of victims. 

When the punishment coverage ratio exceeds about 80\%, the growth trend of cooperation tends to be gentle, indicating diminishing marginal benefits from punishment. Under resource constraints, increasing the spatial density of punishment distribution can enhance overall cooperation at a lower cost.

\begin{figure}  
	\centering  
	\includegraphics [width=\columnwidth]{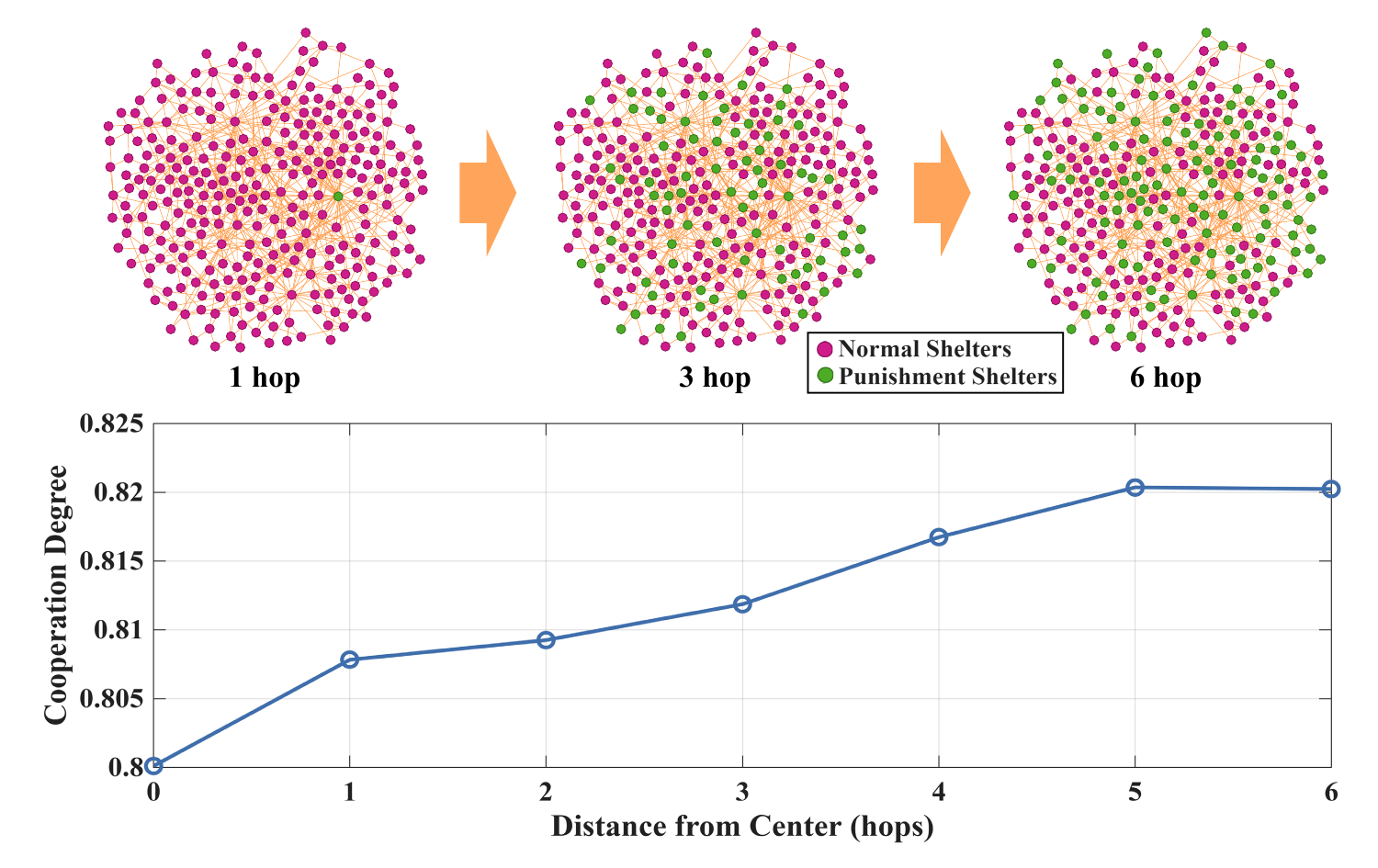}  
	\caption{Targeted punishment from hubs yields efficient cooperation gains.  
	Upper: shelter snapshots under different punishment radii \(h = 1/3/6\). 
    Lower: cooperation degree \(\mu(t)\) cooperation increases with punishment radius until saturation around 5 to 6 hops, the simulation settings unchanged.} 
	\label{7PD} 
\end{figure}

In addition, Fig.~\ref{7PD} indicates that expanding the punishment radius from the central shelter can enhance victims' cooperation and stabilize at $h \ge 5$. Compared with Fig.~\ref{6PR}, this structure-based deployment improves overall cooperation more effectively by leveraging network centrality and achieves a wider spread at a lower cost. Therefore, in a post-disaster environment with limited resources, targeted punishment based on the network structure is more effective in achieving higher cooperation efficiency at lower intervention costs. 

In summary, combined punishment measures effectively suppress defection, while directional management based on network centrality enhances cooperation at lower intervention costs through cross-layer feedback and strategy imitation. 

\subsection{The impact of network coupling on cooperation} 
\begin{figure}  
	\centering  
	\includegraphics[width=\columnwidth]{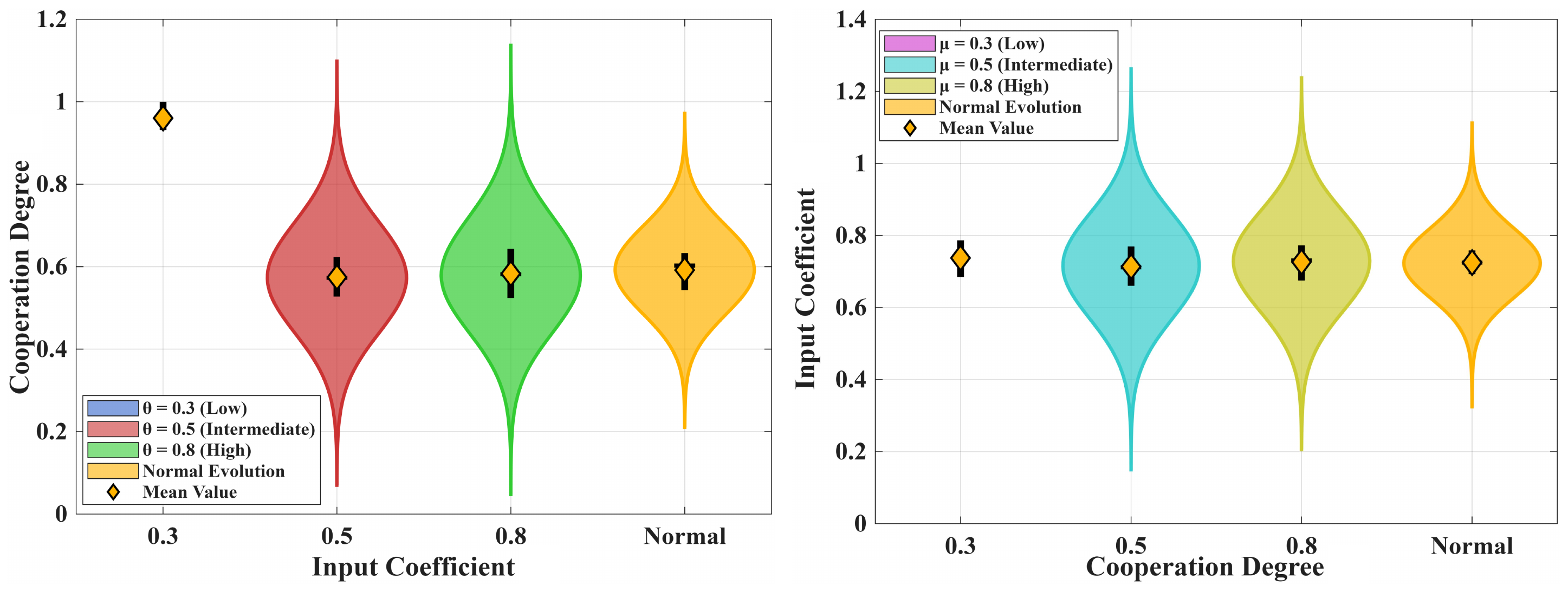}  
	\caption{Shelter cooperation initially suppresses, then promotes victim cooperation, exerting a stronger influence than the reverse effect. Left: effect of shelter input on victim cooperation (violin plot showing distribution, median, mean). Right: effect of victim cooperation on shelter input, displaying a weaker and more symmetric response. Default parameters, the simulation settings unchanged.}
	\label{SVVS}  
\end{figure}

To investigate network coupling, this study examines how cooperation in one layer evolves in response to fixed mechanisms and strategies in the other by independently controlling parameters and behaviors of shelters and victims.
As shown in Fig.~\ref{SVVS}, victim cooperation first decreases, then increases with rising shelter input. When shelter cooperation is low, victims must rely on mutual aid to sustain payoff and achieve high cooperation. At medium level (\(\theta=0.5\)), the increased utility in the shelters weakens the need for mutual assistance among victims. At high input levels, sustained punishment and strong inter-shelter supervision suppress defection, enabling cooperative strategies to prevail and restoring a stable high-cooperation state.

Correspondingly, when victim cooperation is low ($\mu = 0.3$), shelters increase their input to maintain stability, but as victim cooperation stabilizes at medium levels, their input motivation declines. At high levels of victim cooperation $(\mu = 0.9)$, punishment weakens, while subsidies strengthen. As the shelters' income structure becomes more stable, the strategy distribution becomes more concentrated, restoring a stable, high-cooperation state.

These results indicate that as cooperation rises in one layer, the other undergoes a two-phase shift: an initial decline followed by recovery. The upper layer has a more direct and dominant influence, while the feedback of the lower layer is more moderate and delayed, reflecting the asymmetric coupling characteristics of the two-layer network. 

Overall, the stability of post-disaster cooperation depends on a balance between upper-layer incentives and lower-layer constraints, which fosters sustained coordination through feedback and strategy imitation.

\section{Case Analysis}
To verify the applicability and practicality of the proposed post-disaster emergency resource redistribution model and mechanisms, an empirical analysis is conducted based on the actual coordinates of shelters in Beijing. By adjusting key parameters, this study examines the dynamic evolution of cooperation within the two-layer network and compares the results with those obtained on scale-free (BA) networks. 

\begin{figure} 
	\centering  
	\includegraphics [width=\columnwidth]{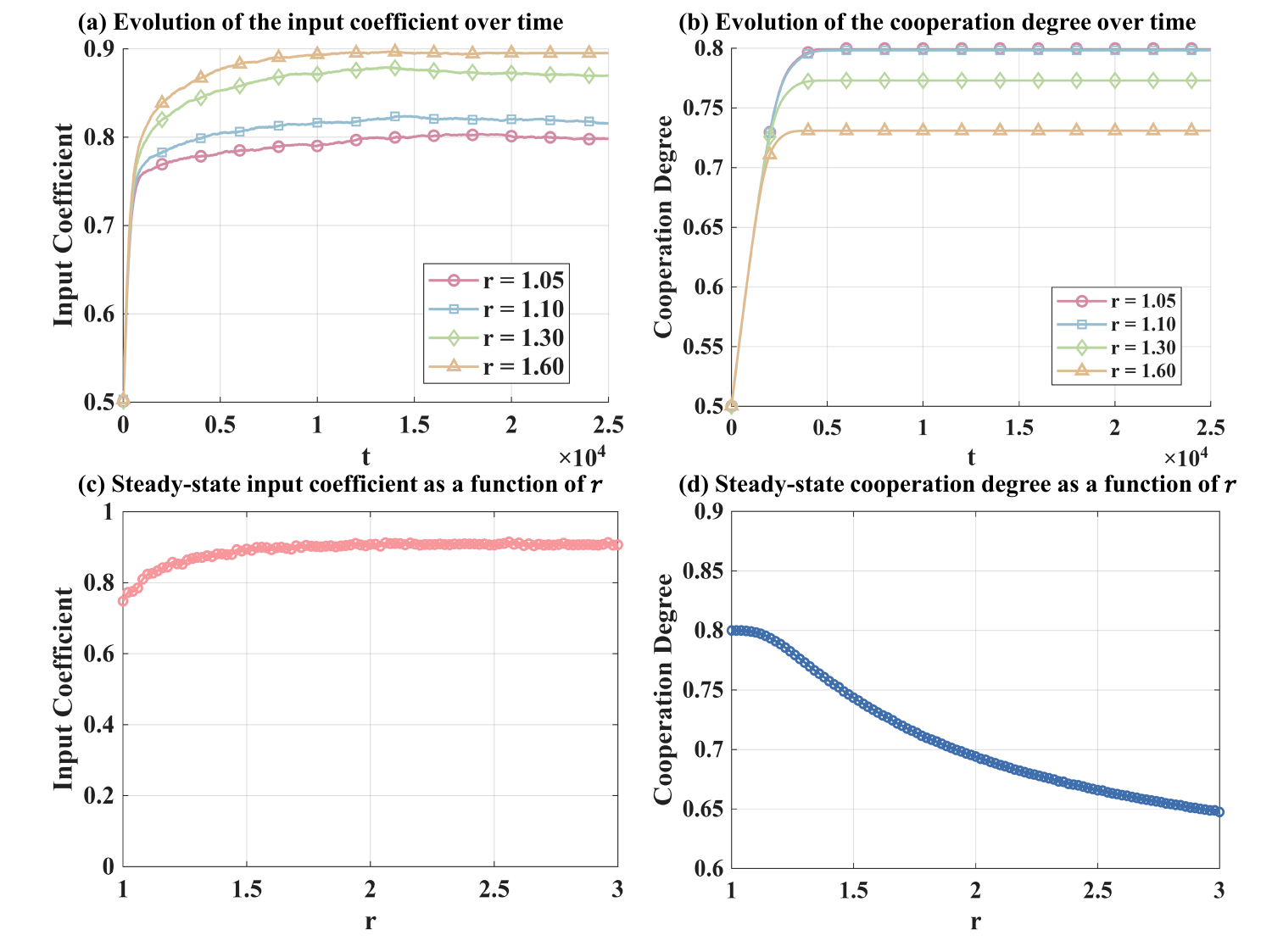}  
	\caption{Moderate enhancement factor \(r\) promotes shelter input, whereas excessive \(r\) weakens incentives, leading to declining cooperation. (a)–(b) Dynamic evolution of $\theta(t)$ and $\mu(t)$. (c)–(d) Steady-state trends matching Fig.~\ref{BA_r} but with more pronounced decline in $\mu(t)$.
	The simulations are conducted on the BJ network, with settings and parameters consistent with those in Fig.~\ref{BA_r}.}  
	\label{BJ_r}  
\end{figure}

The results in Fig.~\ref{BJ_r} indicate that as $r$ increases, shelter input first rises then falls, while victim cooperation steadily declines due to weakened incentives and increased free-riding.

\begin{figure} 
	\centering  
	\includegraphics [width=\columnwidth]{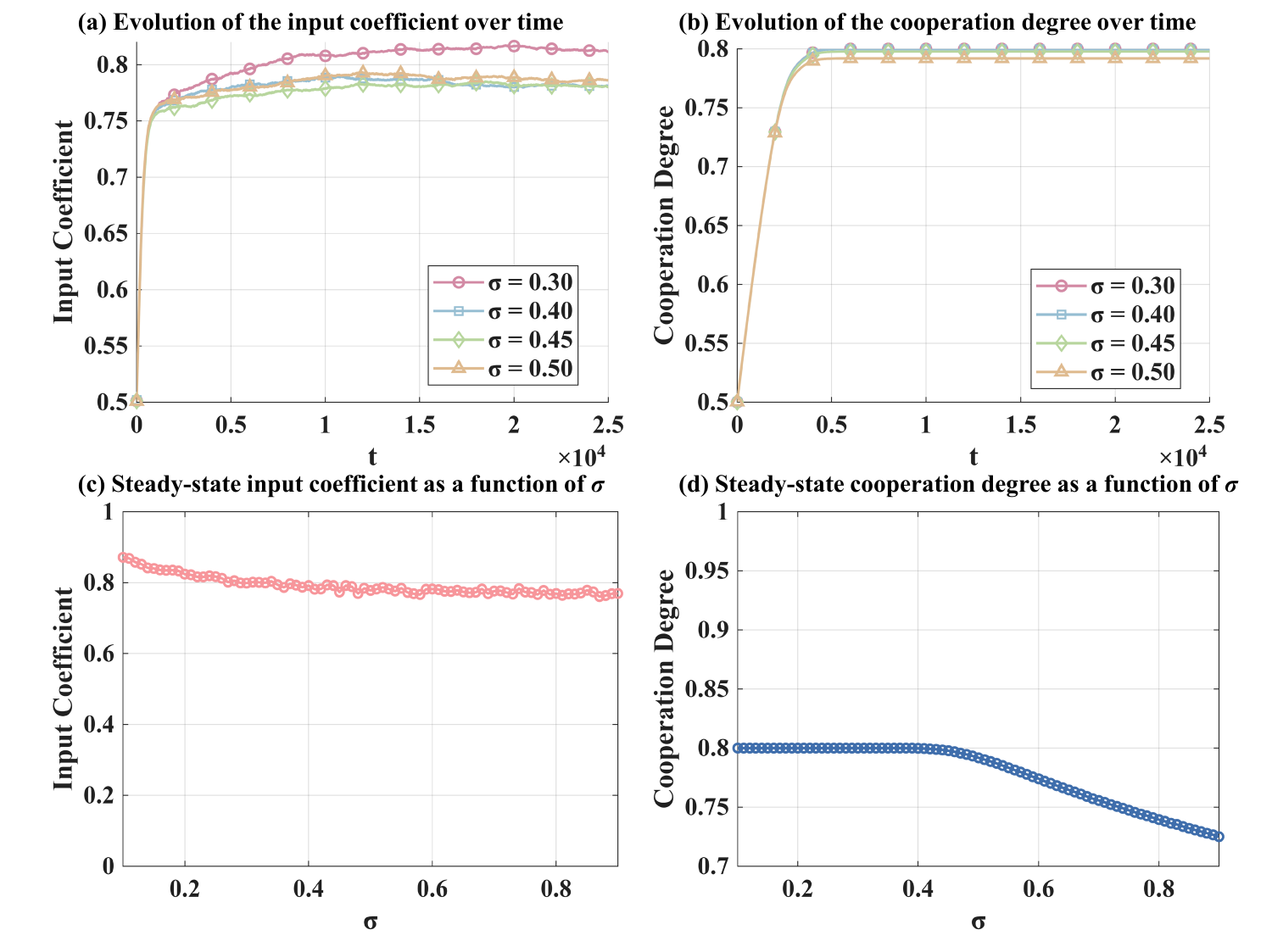}  
	\caption{Subsidies $\sigma$ slightly promote upper-layer cooperation within a threshold but suppress both layers under excessive intensity. (a)–(b) Dynamic evolution of $\theta(t)$ and $\mu(t)$. (c)–(d) Steady-state results showing a monotonic decline of $\mu(t)$ with $\sigma$. The simulations are conducted on the BJ network, with settings and parameters consistent with those in Fig.~\ref{BA_sigma}.}  
	\label{BJ_sigma}  
\end{figure}

As shown in Fig.~\ref{BJ_sigma}, the average input coefficient of shelters slightly decreases as the subsidy intensity increases, whereas the cooperation degree of victims declines markedly. This finding indicates that although generous subsidies can temporarily stabilize the payoffs of shelters, they reduce the shelters' motivation to stimulate cooperation among victims, thereby lowering the overall cooperation level of the two-layer system. 

\begin{figure} 
	\centering  
	\includegraphics[width=\columnwidth]{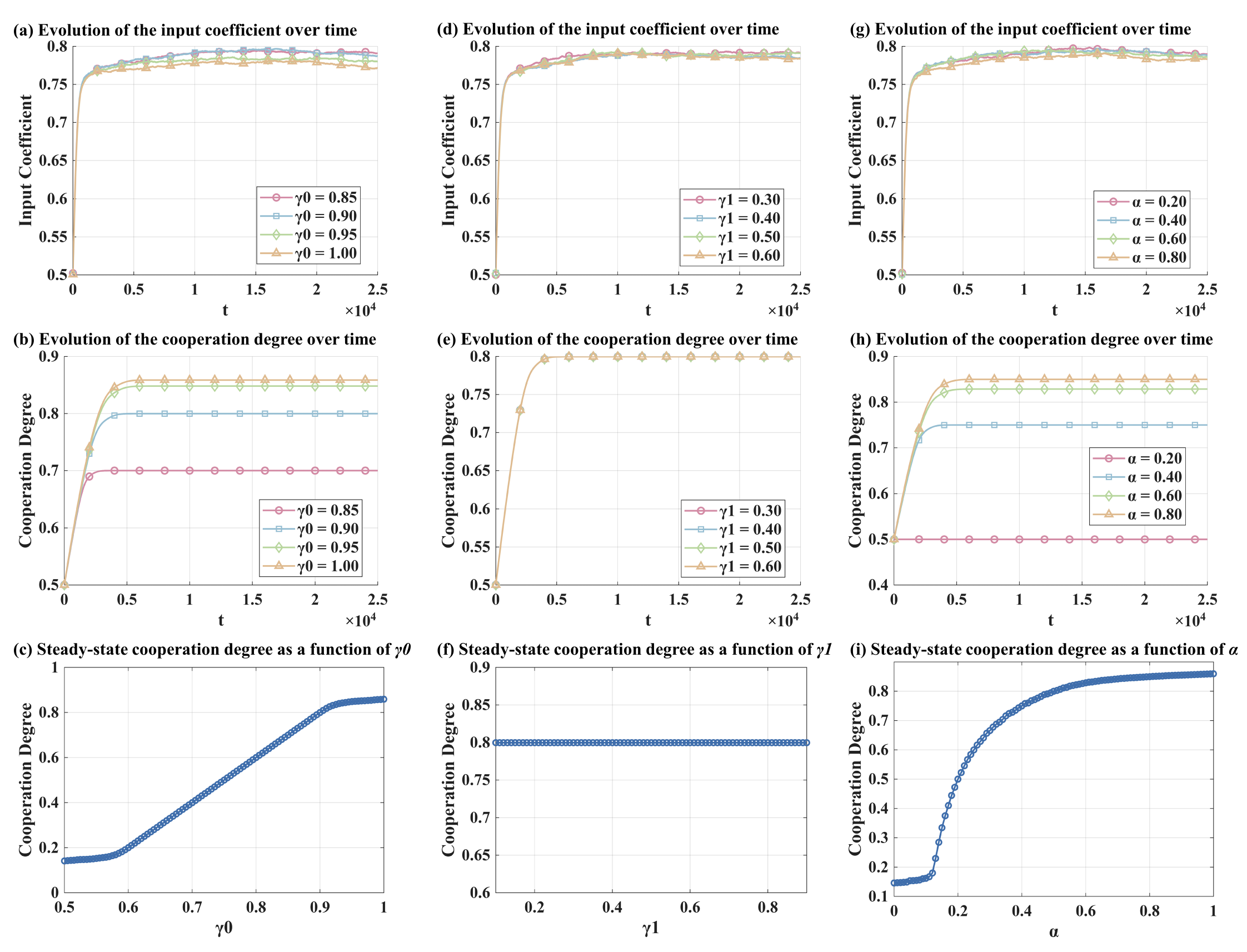}  
	\caption{\(\gamma^0\) serves as the dominant factor, \(\alpha\) plays a secondary role, and \(\gamma_1\) has a minor influence. Rows 1 and 2: dynamics for varying $\gamma_{0}$, $\gamma_{1}$, and $\alpha$. Row 3: steady-state $\mu(t)$ for each parameter, consistent with Fig.~\ref{BA_P}.
	The simulations are conducted on the BJ network.} 
	\label{BJ_P} 
\end{figure}

According to the results in Fig.~\ref{BJ_P}, the coefficient \(\gamma^0\) continues to dominate the promotion of cooperation, \(\alpha\) exerts a secondary positive effect, while \(\gamma_1\) shows almost no influence. 

The results are consistent with those on the BA network, though the effects are slightly weaker. This difference arises because the spatial locality of the BJ network limits the transmission efficiency of both punishment and subsidies, causing local clusters of cooperation to emerge first and gradually diffuse throughout the network.

\begin{figure}  
	\centering  
	\includegraphics [width=\columnwidth]{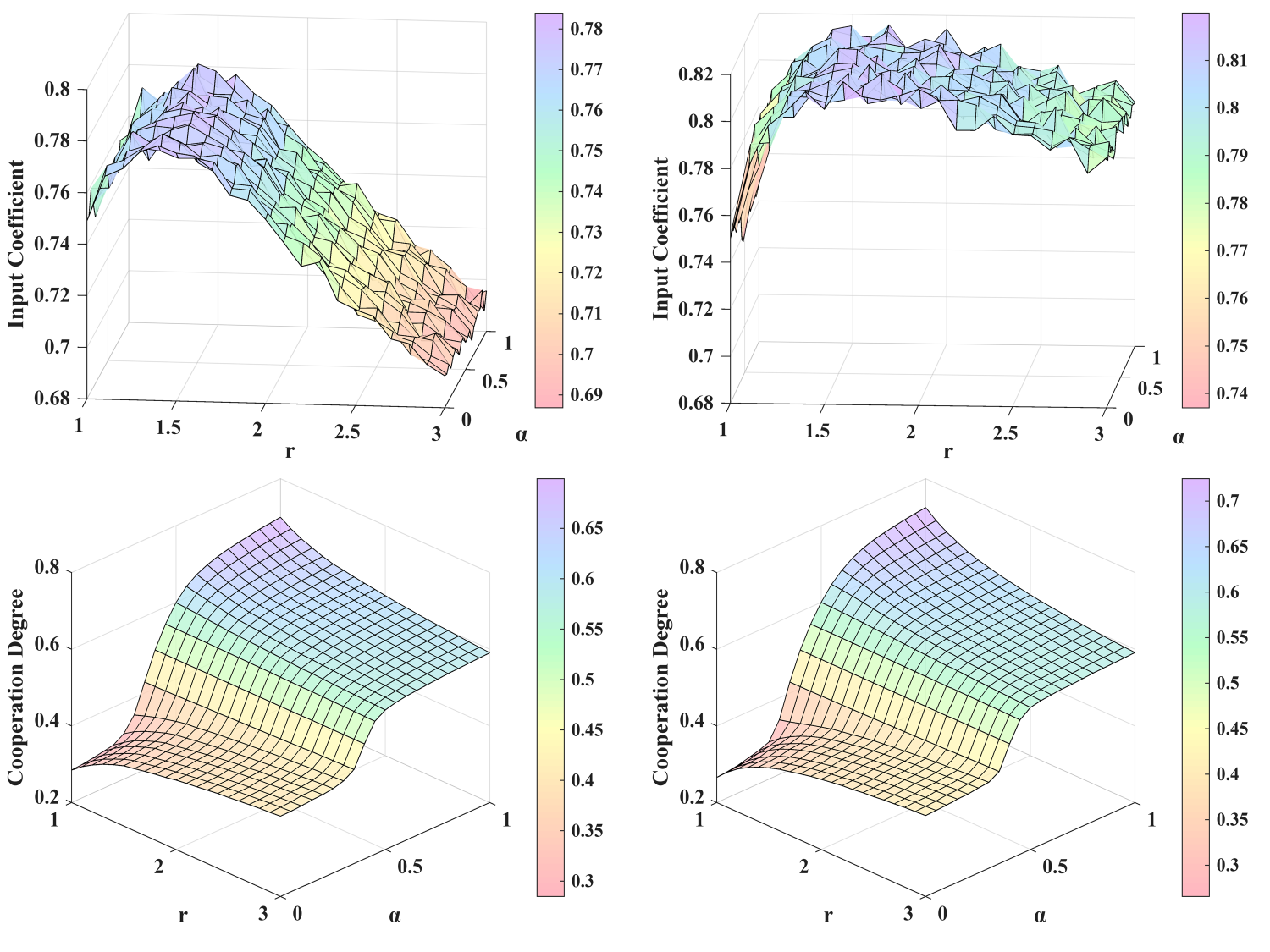} 
	\caption{Effective victim-punishment execution coefficient $\alpha$ amplifies the cooperative impact of moderate enhancement factor $r$, whereas excessive $r$ suppresses cooperation. Each surface shows the steady-state cooperation degree under different ($r$, $\alpha$) combinations on BA (left) and BJ (right) networks. The settings and parameters are consistent with those in Fig.~\ref{BA_r}. } 
	\label{3D}  
\end{figure}

Building on previous findings, Fig.~\ref{3D} further demonstrates that the cooperation patterns exhibit a consistent trend across both network types. When the victim-punishment execution coefficient $\alpha$ is low, increases in $r$ have a limited effect on promoting cooperation. When $\alpha$ exceeds a certain threshold, a pronounced cooperative enhancement region emerges around moderate values of $r$. This indicates that the punishment mechanism amplifies the positive impact of public goods enhancement by improving the efficiency of payoff constraints and strategy imitation. 

Meanwhile, the effect appears slightly weaker in the spatial network generated from the actual coordinates of shelters in Beijing, suggesting that geographical constraints reduce both the range and intensity of cooperation diffusion. Nevertheless, the overall trend remains consistent with that observed in the scale-free network.

\begin{figure}  
	\centering  
	\includegraphics [width=\columnwidth]{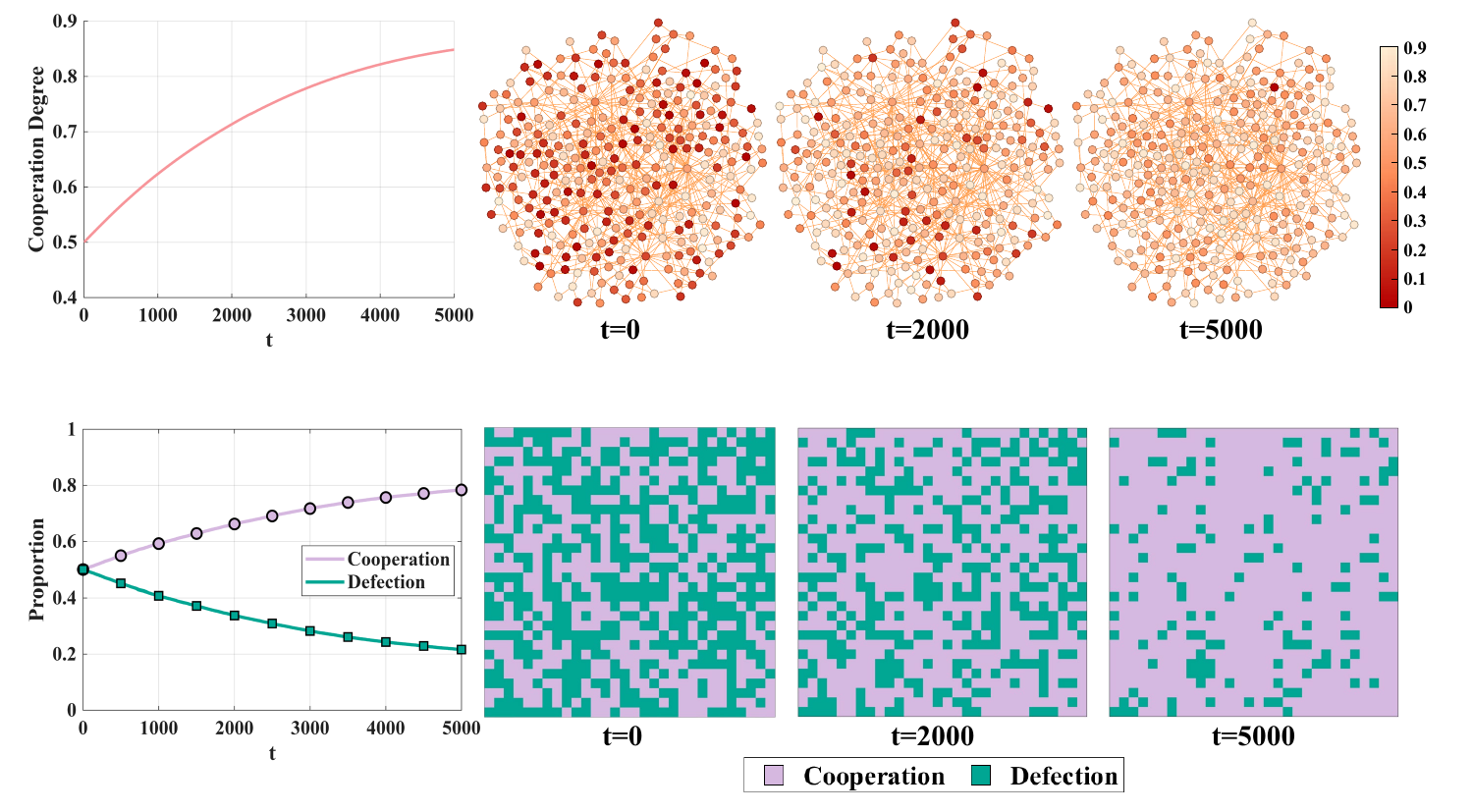}  
	\caption{Cooperation evolution process on the BA network.  
	Upper: temporal evolution of shelter cooperation (\(t = 0,\,2000,\,5000\)), showing cooperation spreading from highly connected nodes.
    Lower: spatial distribution of victim strategies, reflecting the impact of shelter-level dynamics on local interactions. The settings and parameters are consistent with those in Fig.~\ref{BA_r}.}  
	\label{BA_e}  
\end{figure}

Simulation results in Fig.~\ref{BA_e} show that under the combined effects of incentives and punishment, the number of cooperative nodes increases continuously over time, and defectors gradually become marginalized.

\begin{figure}  
	\centering  
	\includegraphics [width=\columnwidth]{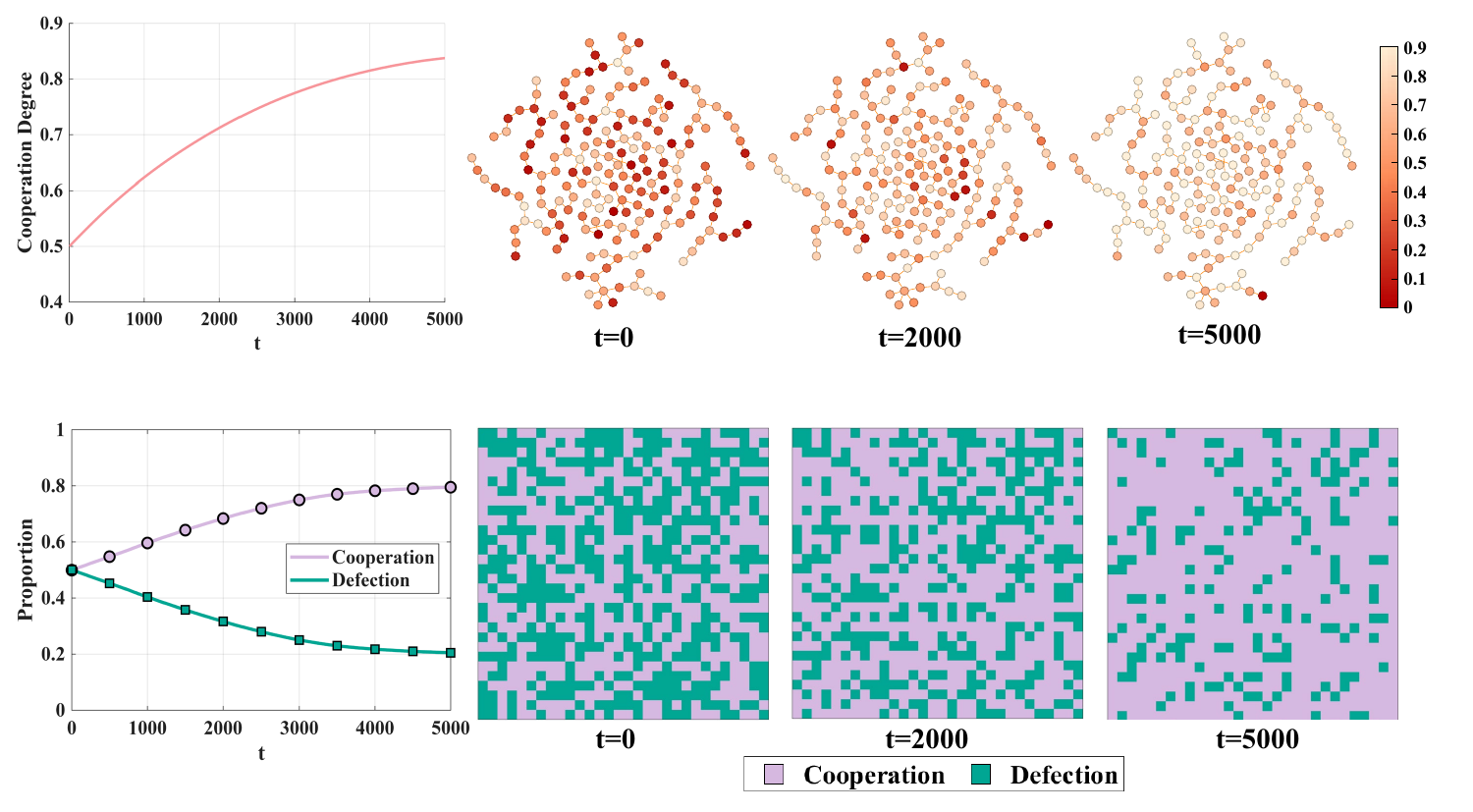}  
	\caption{Cooperation evolution on the BJ network. The dynamics mirror Fig.~\ref{BA_e}, with cooperation spreading network-wide and reflected in victim-level patterns.}     
    \label{BJ_e}  
\end{figure}

The two-layer network evolves from a mixed state to a stable high-cooperation state. This phenomenon reflects how payoff feedback and strategy imitation form a sustained positive feedback loop that maintains synchronized, high-level cooperation between upper-layer shelters and lower-layer victims. Similar to the simulation results above, the results in Fig.~\ref{BJ_e} show that cooperation gradually spreads under the joint influence of incentives and punishments and eventually forms a stable high-cooperation state. However, the growth rate of cooperative nodes is slightly slower in the real network, and certain local areas remain in a mixed state for a longer period. 
 
In summary, the comparative analysis shows that although geographical constraints render network connections more localized and heterogeneous and slightly slow the diffusion of cooperation, the overall evolutionary patterns remain consistent with those observed in scale-free networks. This consistency further verifies the universality and applicability of the proposed model and mechanisms.

\section{Discussion}
This study offers several implications for post-disaster emergency management and resource coordination. For example, moderate and conditional subsidies should be implemented in the early stage rather than indiscriminately increasing subsidy intensity. Subsidies should be tied to cooperation levels and adjusted in stages to avoid ``benefit overflow'' and ``incentive weakening''.  

Second, a reliable punishment system must be established to eliminate the payoff advantage of defection and ensure effective enforcement through a higher execution coefficient. Shelter-level punishment should serve as a supplementary mechanism to maintain post-disaster order.  

Third, under limited resources, network-centrality-based targeted management achieves higher cooperation diffusion efficiency than random deployment. Since upper-layer policies typically take effect earlier than lower-layer responses, sustained incentive and feedback loops are needed to shorten the cooperation recovery cycle.  

Moreover, due to geographic constraints, peripheral shelters often respond more slowly to policy signals, leading to delayed cooperation formation. Therefore, short-term intensive subsidies should be replaced with sustained incentive--punishment coupling to accelerate the stabilization of cooperative behavior.  

\section{Conclusion}
This study developed a coupled two-layer network model that links shelter-level emergency resource redistribution with victims' cooperative behavior. Using simulations on scale-free networks and a real-world Beijing network, the analysis examines how incentive and punishment mechanisms shape cooperation through cross-layer feedback.

The proposed framework captures asymmetric cross-level interactions, with integrates continuous rules on upper-layer and discrete behavioral rules on lower-layer. The results demonstrate that moderate enhancement factors and subsidies promote cooperation, whereas excessive incentives reduce shelter participation and induce free-riding. Consistently enforced punishment effectively suppresses defection and stabilizes cooperation, particularly during early response stages when behavioral uncertainty is high. Targeted interventions on hub shelters improve cooperation diffusion efficiency under resource constraints. An asymmetric cross-layer influence is also observed: upper-layer decisions propagate rapidly and strongly, while lower-layer behavioral responses are weaker and delayed. The consistency between real-world and synthetic networks confirms the robustness of these mechanisms.

This study represents heterogeneity with limited parameters and resource ranges, without capturing deeper social factors such as behavioral bias, reputation dynamics, or information asymmetry. Second, the model is based on static network simulations, whereas real-world emergency management is dynamic and multi-staged. Future work could incorporate time-varying resources, evolving costs, and real-time data to develop adaptive, topology-changing networks that better reflect post-disaster cooperation and policy responses.

Overall, this study advances the theoretical understanding of cooperation evolution in hierarchical systems and provides practical insights for post-disaster emergency management. By balancing incentives and constraints, improving punishment enforcement, and optimizing spatial deployment, rapid cooperation can emerge within the critical response window, guiding effective resource and personnel coordination.

\subsection*{SUPPLEMENTARY MATERIAL}
See the supplementary material for detailed pseudocode illustrating the cross-layer strategy interaction and updating process of the proposed model. It first provides an overview of the updating workflow, followed by step-by-step pseudocode for each algorithmic component.

\begin{acknowledgments}
This work is supported by the National Key R\&D Program of China (2021YFA1000402),  
the National Natural Science Foundation of China (72571215), 
the Guangdong Basic and Applied Basic Research Foundation (2024A1515011241), the Key Scientific Research Program Project of Shaanxi Provincial Department of Education (24JS055), and the Natural Science Basic Research Plan in Shaanxi Province of China (2024JC-YBMS-588).
\end{acknowledgments}

\section*{AUTHOR DECLARATIONS}

\subsection*{Conflict of Interest}
The authors have no conflicts of interest to disclose.

\subsection*{Author Contributions}
Y.C. designed the model, performed the simulations, and wrote the manuscript.
G.X. supervised the study and provided critical revision of the manuscript.
S.F. assisted with model development and data analysis.
C.W. contributed to result interpretation and manuscript editing.
All authors reviewed and approved the final manuscript.

\section*{DATA AVAILABILITY}
Some or all of the data and models supporting the findings of this study are available from the corresponding author upon reasonable request. The full source code is available from the corresponding author upon reasonable request.

%% \bibliographystyle{aipnum4-1}
%\bibliography{aipsamp.bib}
%merlin.mbs aipnum4-1.bst 2010-07-25 4.21a (PWD, AO, DPC) hacked
%Control: key (0)
%Control: author (8) initials jnrlst
%Control: editor formatted (1) identically to author
%Control: production of article title (-1) disabled
%Control: page (0) single
%Control: year (1) truncated
%Control: production of eprint (0) enabled
\providecommand{\noopsort}[1]{}\providecommand{\singleletter}[1]{#1}%

\end{document}